\documentclass{rmf-d}
\usepackage{nopageno,rmfbib,multicol,times,epsf,amsmath,amssymb,cite}
\usepackage[T1]{fontenc} 
\usepackage[]{caption2}
\usepackage{graphicx}
\usepackage{blindtext}
\usepackage{color}
\usepackage{hyperref}
\def\rmfcornisa{Research OR Education of Physics \hfill\rmf\ {\bf ??} (*?*) ???--???
\hfill MES? A\~NO?}

\def\rmfcintilla{{\it Rev.\ Mex.\ Fis.\/} {\bf ??} (*?*) (????) ???--???}
\clearpage \rmfcaptionstyle 
\begin{document}
\markboth{ RMF Editorial Team    }{ A \LaTeX template for the RMF, RMF-E, SRMF }

%
%
\title{  Top quark and vector meson production in heavy ion collisions at CMS
\vspace{-6pt}}
\author{  Luis F. Alcerro
\thanks{Presented at ``XVIII Mexican Workshop on Particles and Fields'', Puebla, Mexico, November 21-25, 2022.}
\thanks{Supported by the Nuclear Physics program\href{https://pamspublic.science.energy.gov/WebPAMSExternal/Interface/Common/ViewPublicAbstract.aspx?rv=00d4fe0f-48a0-4d4a-baf1-c70867d9e499&rtc=24&PRoleId=10}{ DE-FG02-96ER40981} of the U.S. Department of Energy.}
}

\address{   Department of Physics \& Astronomy, The University of Kansas, Malott Hall, room 1082
1251 Wescoe Hall Dr.
Lawrence, KS 66045, US    }
%
%
\author{ }
\address{ }
\author{ }
\address{ }
\author{ }
\address{ }
\author{ }
\address{ }
\author{ }
\address{ }
\maketitle
%
%
\begin{abstract}
\vspace{1em} 
%
%
Heavy ion collisions are rich and complex systems that involve different aspects of QCD and electromagnetic phenomena. From head-on collisions to the case in which the nuclei miss each other, many of QCD and photo-induced probes are being actively investigated. 
In this contribution we briefly discuss both of these different topics and some of the CMS results on top quark production and vector meson photo-production, as well as the physics motivation to study these processes are presented. 
\vspace{1em}
\end{abstract}
\keys{ \bf{\textit{ Nucleus collisions, QGP, UPCs, Top quark, Upsilon
}} \vspace{-8pt}}
\begin{multicols}{2}

\section{Introduction}

The Compact Muon Soleniod experiment (CMS) at the Large Hadron Collider (LHC) has a vibrant heavy ion collision program, studying a wide variety of particle and nuclear physics phenomena, ranging from perturbative QCD to quark matter to electromagnetic processes. Paradoxically, heavy ion collisions reproduce the messiest and cleanest events at LHC in central and ultra-peripheral collisions (UPC), respectively. 

Central collisions of heavy ions are used to create matter under extreme conditions of temperature and energy density, known as Quark-Gluon Plasma (QGP). Many questions related to QGP properties remain elusive, some of which are being addressed at different LHC experiments. In particular, the multi-TeV energies reached at the LHC have opened the opportunity to measure top quarks in heavy ion collisions, providing \textemdash for the first time \textemdash the possibility to investigate the time evolution of the QGP. 

On the other hand, UPCs offer the opportunity to study photo-induced processes, such as vector meson photo-production. These are very clean processes and powerful probes to understand the gluonic structure of nuclei.  

\section{Top quark}
 The top quark \textemdash with a mass of roughly $173$ GeV \textemdash is the heaviest particle in the Standard Model and it is mainly produced at LHC in top-antitop ($\mathrm{t\bar{t}}$) pairs by gluon fusion. It decays most of the time into a b quark and a W boson, the latest eventually decaying into either leptons or quark-antiquark $(\mathrm{q\bar{q}})$ pairs. 

The CMS Collaboration presented in 2020 the very first evidence of top quark in heavy ion collisions using data recorded in 2018, corresponding to an integrated luminosity of 1.7 nb$^{-1}$ of lead-lead collisions at center-of-mass energy of $\sqrt{s_\mathrm{NN}}=5.02$ TeV \cite{CMS:2020aem}. The dilepton channel \textemdash in which both of the W bosons coming from the top quarks, decay into muons or electrons \textemdash is used in this analysis with and without inclusion of information coming from b tagged jets, referred as dileptons only ($2\ell_{\text{OS}}$) and dileptons + b-jets ($2\ell_{\text{OS}}$ + b-jets) methods, respectively. 

\begin{figure}[H]
 \includegraphics[width=\linewidth]
    {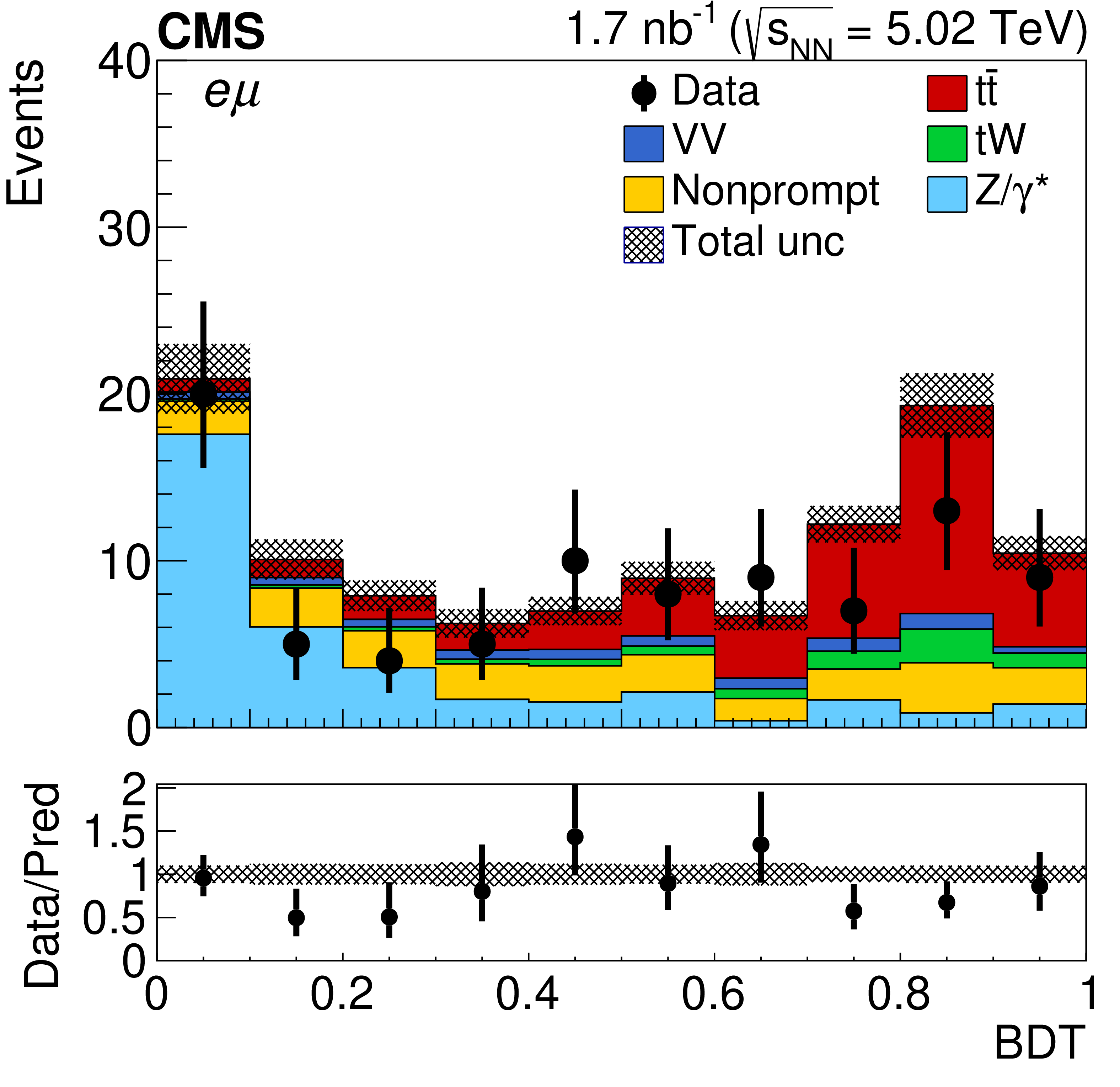}
 \includegraphics[width=\linewidth]
    {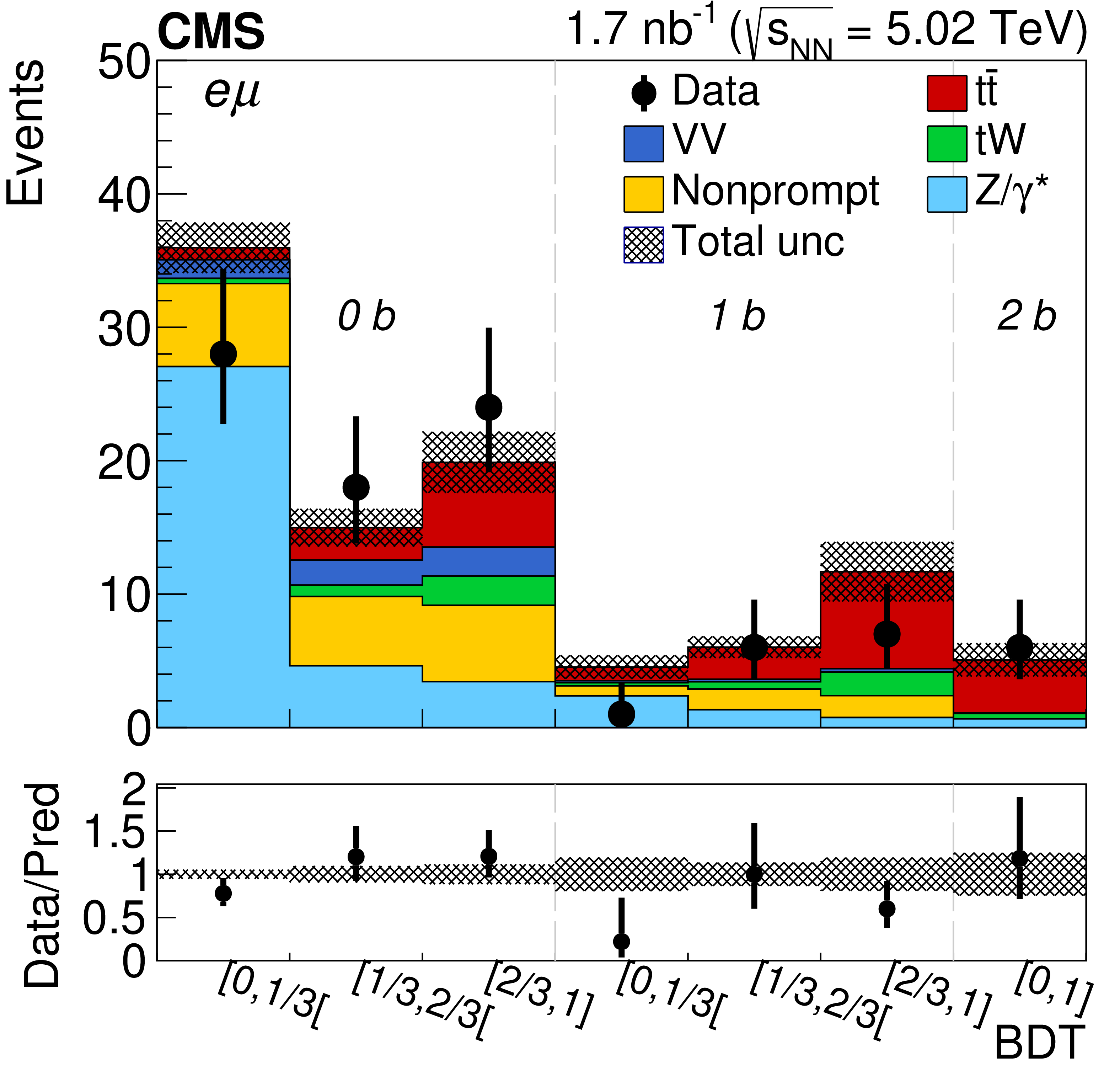}
 \caption{The BDT distributions in the $e \mu$ channel with the dileptons only (dileptons + b-jets) method is shown in the top (bottom) panel \cite{CMS:2020aem}.  }
 \label{fig:bdt1}
\end{figure}

The data is filtered to contain two opposite sign (OS) leptons with $p_\mathrm{T}>25$ (20) GeV and $|\eta|<2.1$ (2.4) for electrons (muons), isolated from hadronic activity. In addition, the presence of b-tagged jets (jets initiated by b quark decays) is required in a second method. In both approaches, Boosted Decision Trees (BDTs) are used to separate signal events from background processes (mainly Drell-Yan, W+jets and QCD multijets), as is shown in Fig. \ref{fig:bdt1}.

\begin{figure}[H]
 \includegraphics[width=\linewidth]
    {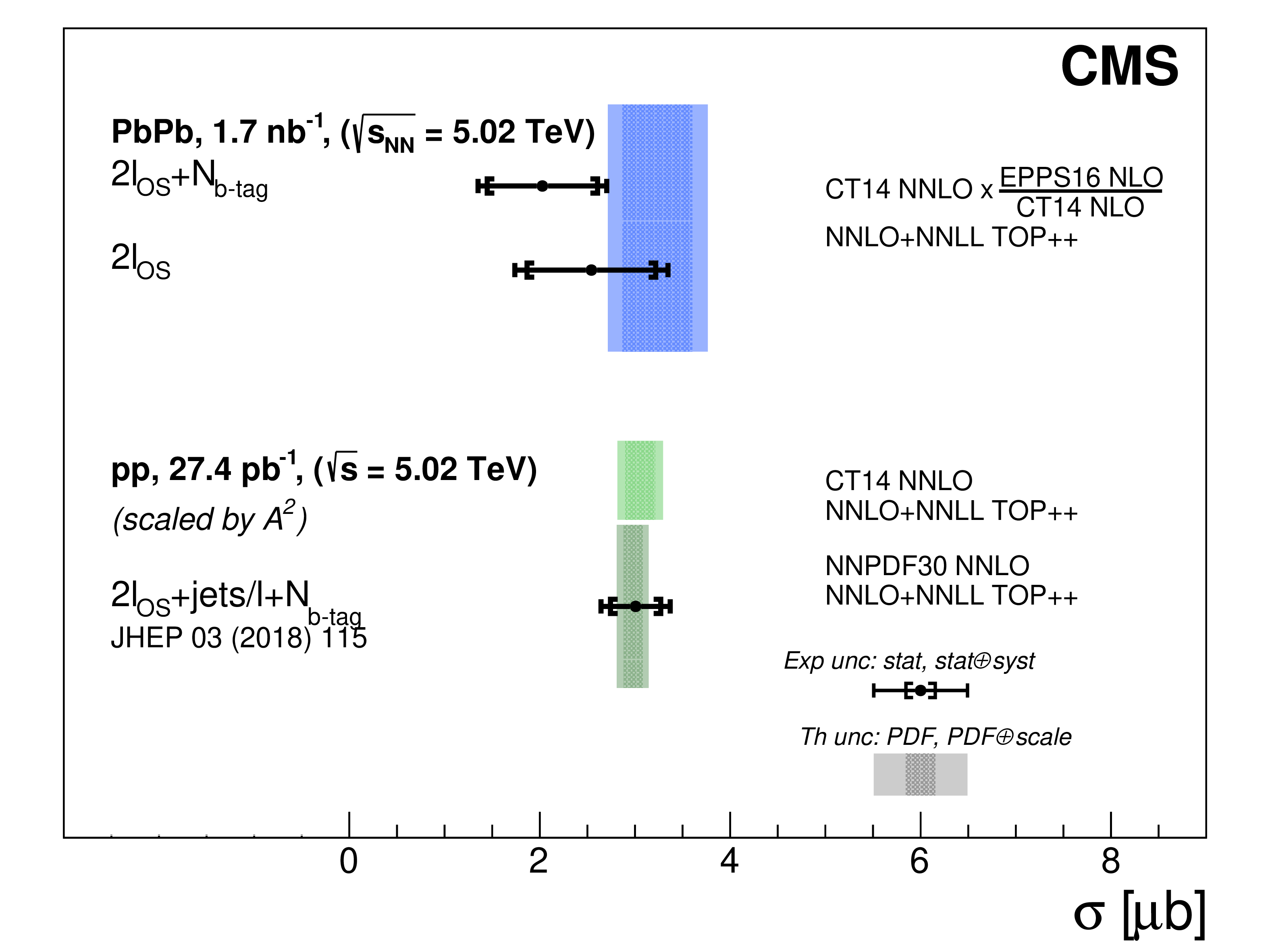}
 \caption{Inclusive $\mathrm{t\bar{t}}$ cross sections measured with both methods in the combined $ee$, $\mu \mu$, and $e \mu$ final states in PbPb collisions at $\sqrt{s}$= 5.02 TeV, and pp results at the same energy \cite{CMS:2020aem}.  }
 \label{fig:xsec1}
\end{figure}

Likelihood fits to binned BDT distributions are performed separately for the two methods to extract the cross section, obtaining:
\[\sigma_{\mathrm{t\overline{t}}} = 2.03^{+0.71}_{-0.64} \text{ }\mu \text{b}\]
with the $2\ell_{\text{OS}} +$ b-jets method and 
\[\sigma_{\mathrm{t\overline{t}}} = 2.54^{+0.84}_{-0.74} \text{ }\mu \text{b}\]
with $2\ell_{\text{OS}}$. The results are compatible with theoretical calculations and pp at 5 TeV scaled by the number of binary nucleon-nucleon collisions in PbPb as well, as is shown in Fig. \ref{fig:xsec1}.

This result is a first step of a novel research line to unveil the time dependence of the QGP. In contrast to all the probes used so far in the literature \textemdash which are sensitive only to time-integrated properties of the QCD medium \textemdash, the top quark represents a promising tool to access the time evolution of the QGP. As a consequence of a series of time delays between the moment of the collision and that when the products of hadronically decaying W bosons produced from top quark decays start interacting with the medium, $\mathrm{t\bar{t}}$ events in heavy nuclei collisions have the potential to bring valuable information of the QGP time structure, as is pointed out in a feasibility study \cite{Apolinario:2017sob}.

\begin{figure}[H]
\includegraphics[width=\linewidth]
    {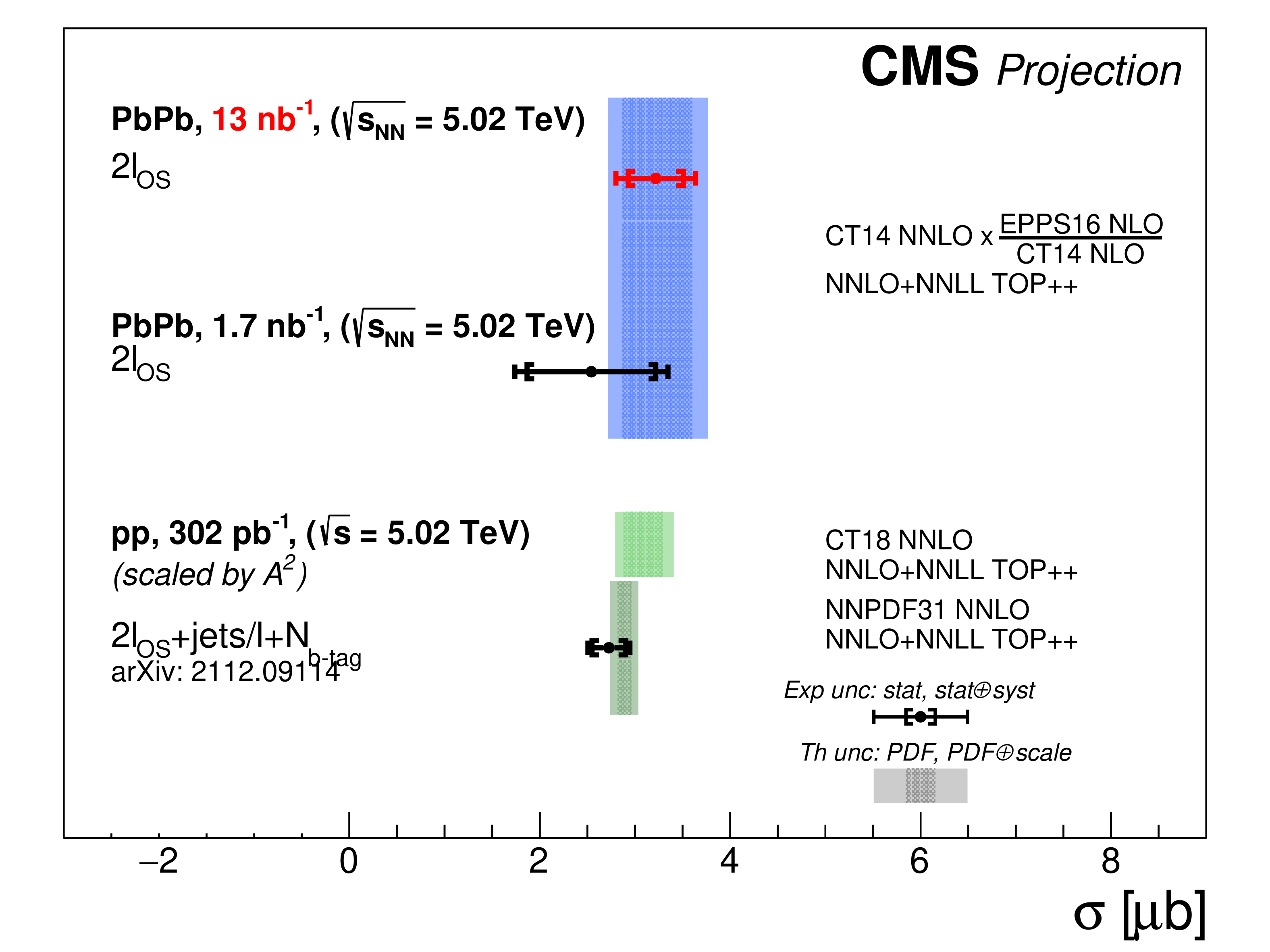}
 \caption{Inclusive cross sections projection for $\mathrm{t\bar{t}}$ in PbPb at HL-LHC \cite{CMS:2020aem}.  }
 \label{fig:xsec2}
\end{figure}

Projection of $\mathrm{t\bar{t}}$ in the context of the High-Luminosity LHC assuming a PbPb integrated luminosity of 13 nb$^{-1}$ is shown in Fig. \ref{fig:xsec2}. This projection (red) takes into account only dilepton final states (no b jets) and is compared with the measured (black) $\mathrm{t\bar{t}}$ cross section, scaled pp data, and theoretical predictions at NNLO+NNLL accuracy in QCD.

\section{Vector meson production in UPCs}

In contrast to central collisions which are mostly devoted to study QGP properties, UPCs are very clean events which interplay QED, QCD and beyond the Standard Model physics. In UPCs, the two heavy ions pass each other with an impact parameter greater than the sum of their radii and the electromagnetic fields around them are highly contracted due to the Lorentz boost, hence these electromagnetic fields can be treated as polarized quasi-real photons. One of these photons produced by the projectile nucleus may fluctuate into a quark-antiquark pair and interact with the target nucleus via gluon exchanges, producing exclusively a vector meson (VM).

\begin{figure}[H]
\includegraphics[width=\linewidth]
    {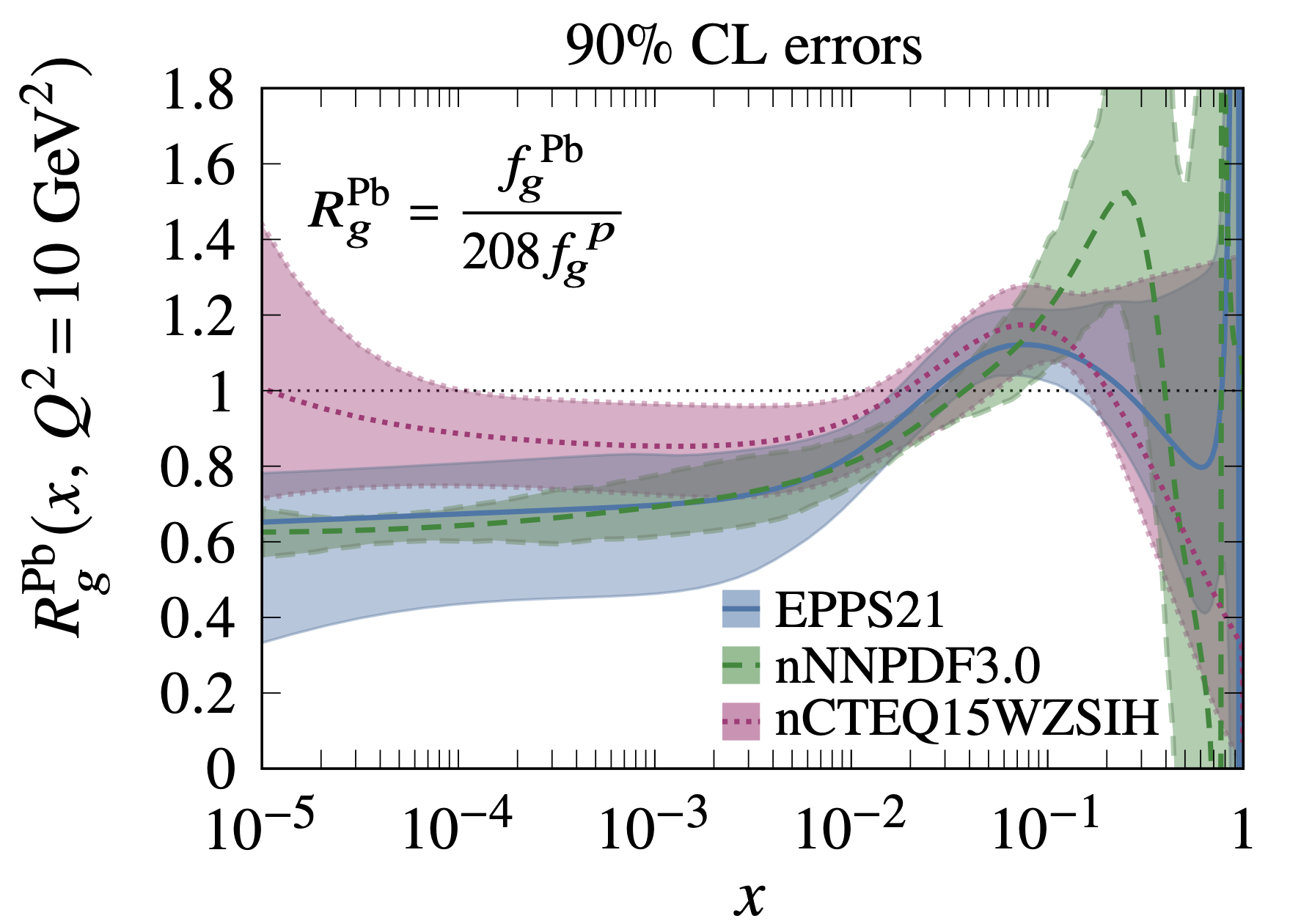}
 \caption{Nuclear modifications of the gluon PDFs in Pb from different nPDF analyses \cite{Paakkinen:2022qxn}. }
 \label{fig:npdf}
\end{figure}

In particular, the coherent VM photo-production \textemdash when the photon interacts with the target as a whole \textemdash in UPCs is relevant to study the gluonic structure of nuclei, since the cross section is proportional at leading order to the gluon Parton Distribution Function  of the target (nPDF). Moreover, heavy-flavor VMs provide access to the poorly known low Bjorken-x region ($10^{-5}$ - $10^{-2}$) at LHC energies, as shown in Fig. \ref{fig:npdf}, where we can notice big uncertainties in these particular region for all shown nuclear PDFs. 

\begin{figure}[H]
\includegraphics[width=\linewidth]
    {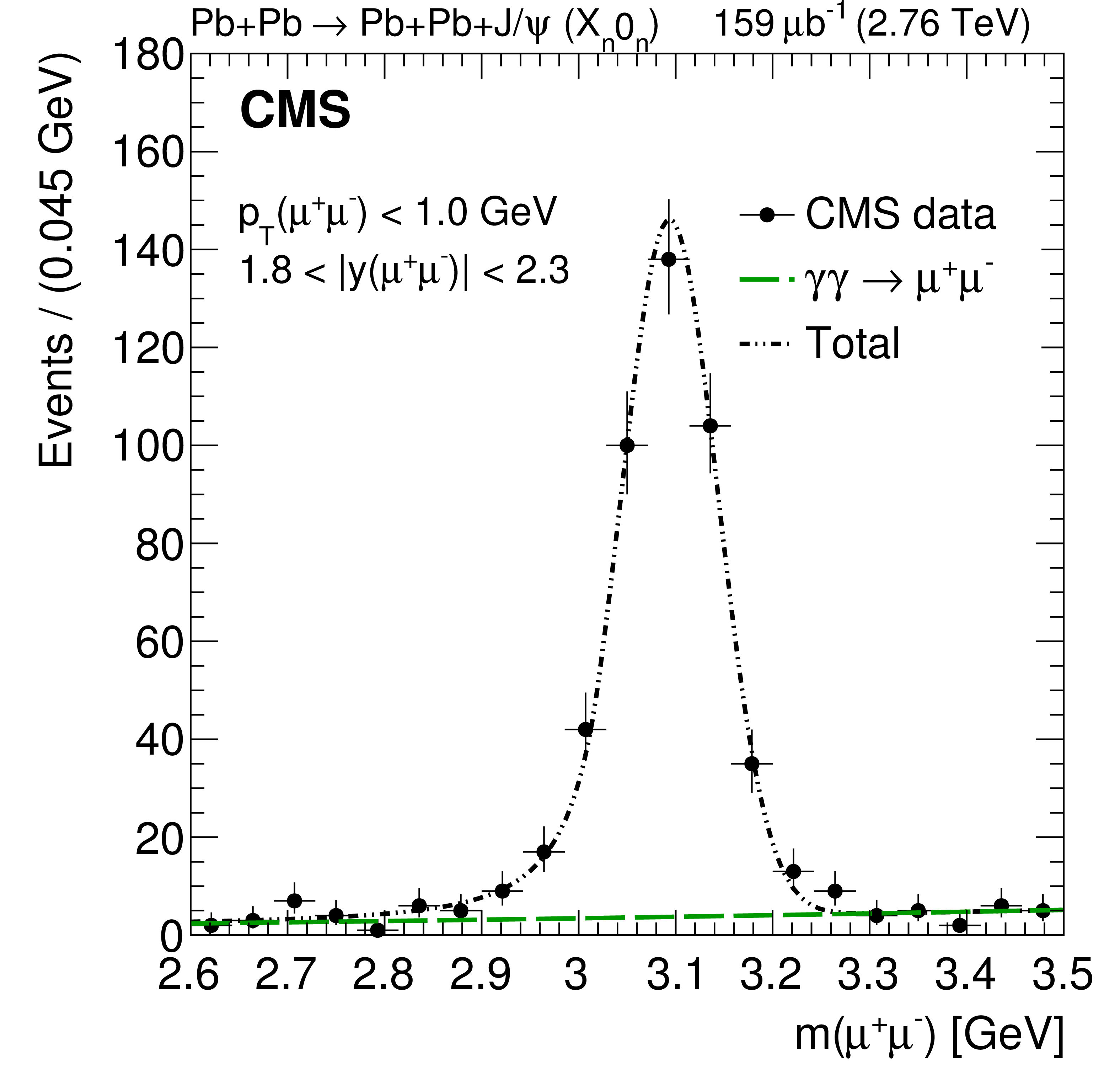}
 \caption{Dimuon invariant mass fit to the data\cite{CMS:2016itn}.}
 \label{fig:invmass}
\end{figure}

Related results have been reported by the CMS Collaboration such as  exclusive photo-production of $\Upsilon$ in pPb at $\sqrt{s_{\mathrm{NN}}}=$ 5.02 TeV \cite{CMS:2018bbk}, $J/\psi$ in PbPb at  $\sqrt{s_{\mathrm{NN}}}=$ 2.76 TeV \cite{CMS:2016itn} and recently at $\sqrt{s_{\mathrm{NN}}}=$ 5.02 TeV
\cite{CMS:2023snh}. In general terms, the analysis strategy relies on the identification of pairs of oppositely-charged muons in an otherwise empty detector within a corresponding dimuon invariant mass ($\mathrm{m}_{\mu \mu}$) window, as shown in Fig. \ref{fig:invmass} for the $J/\psi$ CMS analysis at 2.76 TeV \cite{CMS:2016itn}, where a peak is clearly visible around $\mathrm{m}_{\mu \mu}=$ 3.1 GeV corresponding to the $J/\psi$ mass. Further exclusive selection criteria are also typically applied to the data using forward calorimeters. 

The total measured VM production cross section of a given rapidity is given by
\end{multicols}
\begin{wequation}
    \frac{d \sigma_{AA \rightarrow A A' VM}(y)}{d y } = N_{\gamma / A}(y) \sigma_{\gamma/A \rightarrow VM A'}(y)  + N_{\gamma / A}(-y) \sigma_{\gamma/A \rightarrow VM A'}(-y),
\label{eq:xsec}
\end{wequation}
\begin{multicols}{2}
where $N_{\gamma/ A}$ is the photon flux, $y$ is the VM rapidity, and $\sigma_{\gamma/A \rightarrow VM A'}$ is the VM photo-production cross-section. The presence of two terms in Eq. \ref{eq:xsec} is due to the fact the either nuclei can produce a photon or can serve as a target. The separation of these terms is in fact one of the main difficulties to extract $\sigma_{\gamma/A \rightarrow VM A'}$. A novel solution for this problem has been proposed in \cite{Guzey:2013jaa,Guzey:2016piu} in which the main idea is to control the collision impact parameter by detecting forward neutrons emitted via electromagnetic dissociation (EMD) of ions and classifying events in forward neutron multiplicity categories. In this way, we would have one equation \ref{eq:xsec} corresponding to each neutron category and \textemdash since the photon fluxes can be computed from simulation \textemdash the problem reduces to a simple system of equations.  

\begin{figure}[H]
\includegraphics[width=\linewidth]
    {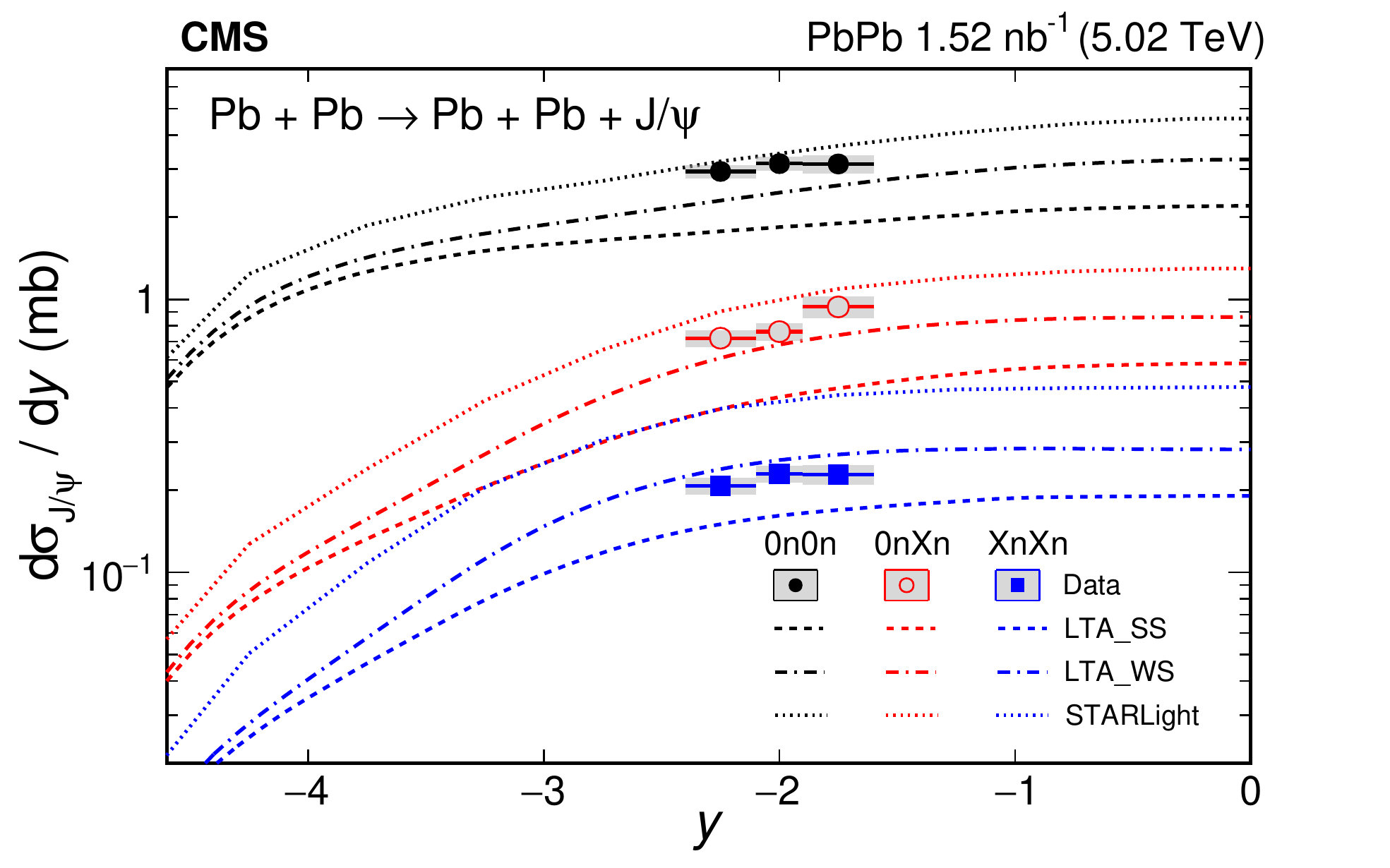}
 \caption{Differential coherent $J/\psi$ photo-production cross section as a function of rapidity in the three different neutron multiplicity classes \cite{CMS:2023snh}.}
 \label{fig:jpsixsec}
\end{figure}

The CMS Collaboration recently reported $J/\psi$ photo-production in PbPb at $\sqrt{s_{\mathrm{NN}}}=$ 5.02 TeV \cite{CMS:2023snh} implementing the aforementioned technique by using the zero-degree calorimeters (ZDC) to tag forward neutrons produced via EMD. Events are classified in classes of forward-neutron multiplicity \textemdash by using the energy depositions in the ZDCs \textemdash as having no neutrons (0n) or with at least one neutron (Xn, X $\geq$ 1), so three neutron multiplicity classes can be identified, namely 0n0n, 0nXn, XnXn. The measured $J/\psi$ photo-production differential cross sections are obtained in the three neutron multiplicity classes, as illustrated in Fig. \ref{fig:jpsixsec}. 

\begin{figure}[H]
\includegraphics[width=\linewidth]
    {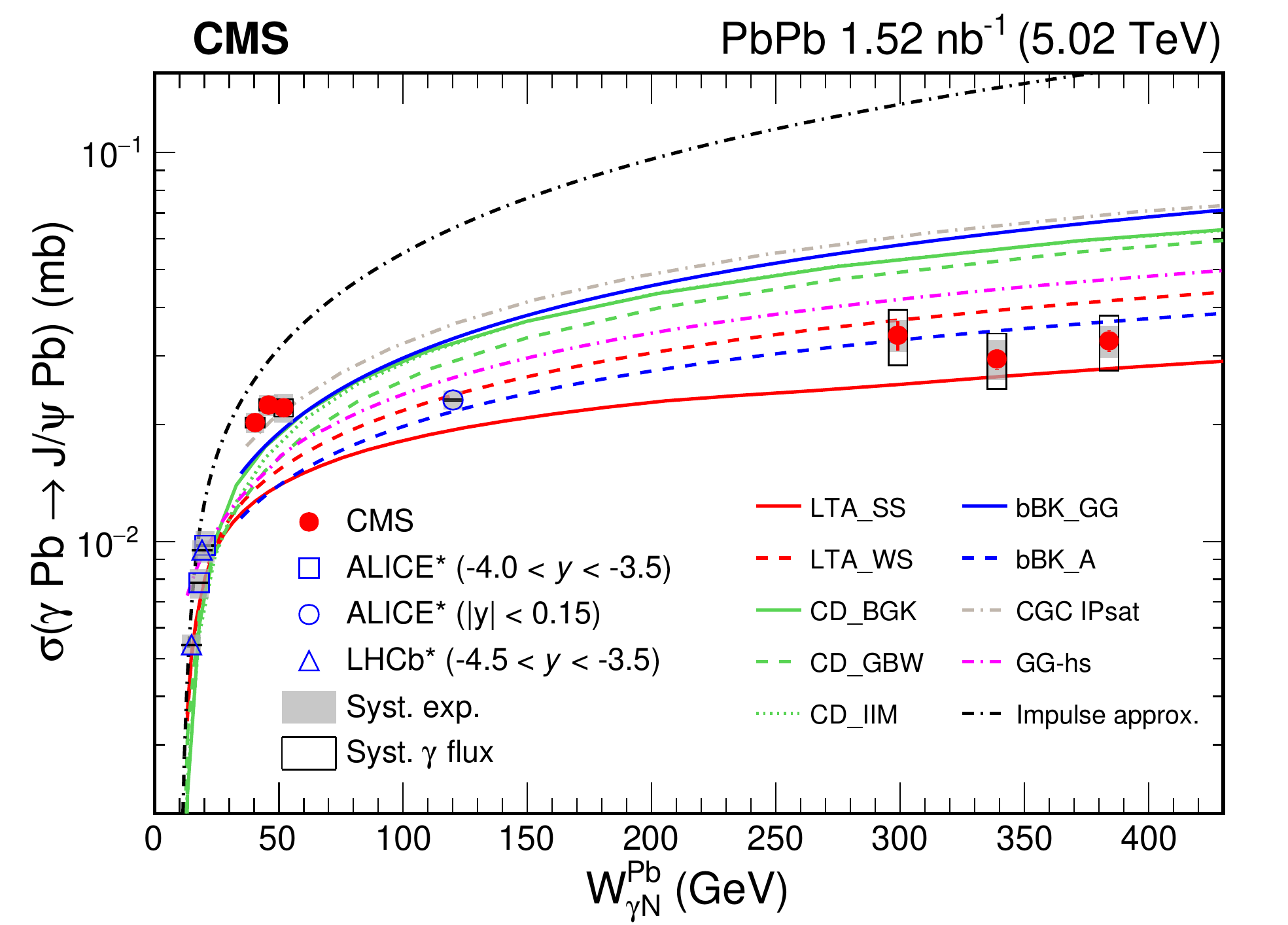}
 \caption{Coherent $J/\psi$ photo-production cross section as a function of $W_{\gamma \mathrm{N}}^{\mathrm{A}}$ after separating the low and high energy photon contributions. ALICE and LHCb measurements as well as theoretical preditions are also displayed \cite{CMS:2023snh}.}
 \label{fig:jpsixsec2}
\end{figure}

After separating the low and high energy photon contributions \textemdash the two terms in Eq. \ref{eq:xsec}\textemdash, the $J/\psi$ photo-production cross section is obtained as a function of the photon-nucleus center-of-mass energy per nucleon $W_{\gamma \mathrm{N}}^{\mathrm{A}}$ and shown in Fig. \ref{fig:jpsixsec2}, together with ALICE and LHCb measurements and theoretical models. As we can notice, the cross section quickly grows as we go from $W_{\gamma \mathrm{N}}^{\mathrm{A}} \approx 15$ to 40 GeV, in agreement with a fast-growing gluon density at low $x$. In the region $W_{\gamma \mathrm{N}}^{\mathrm{A}} \approx 40$ GeV, the cross section starts growing slowly up to 400 GeV and none of the theoretical models are consistent with the measurement. This result could imply the first hints of saturation of the gluon density in the Pb nucleus. Another physical interpretation of of this measurement is in the context of the black-disc limit (BDL) \cite{Frankfurt:2001nt}, in which most of of the target Pb nucleus becomes completely absorptive to photons and the cross section approaches the unitarity limit allowed by the geometric nuclear size. 

\section{Summary}
We have exemplified two different kind of studies in heavy ion collisions that are being actively investigated in CMS. On the one hand, we have stated that the top quark represents a novel probe to study the time structure of the QGP, specially at higher luminosity. One the other hand, VM photo-production in UPCs are powerful probes in the context of low $x$ physics and to try to unveil the internal hadronic structure.

\end{multicols}

\medline
\begin{multicols}{2}
%
\nocite{*}
\bibliographystyle{rmf-style}
\bibliography{ref}
%
%
%

%
%
\end{multicols}
\end{document}